\documentclass[aps,twocolumn,prl,superscriptaddress,tightenlines,longbibliography]{revtex4-1}
\usepackage{CJK}
\usepackage{amsfonts}
\usepackage{graphicx}
\usepackage{hyperref}
\usepackage{amsmath}
\usepackage{amssymb}
\usepackage[usenames, dvipsnames]{color}
\usepackage{subfigure}
\usepackage{lipsum}
\usepackage{lineno}

\begin{document}

\title{Two-Photon Bound States in the Continuum: A No-Go Theorem and
Long-Lived Quasi-Bound States}
\author{Yue Chang}
\affiliation{Beijing Academy of Quantum Information Sciences, Beijing
100193, China}
\email{yuechang7@gmail.com}

\begin{abstract}
Two-photon bound states in the continuum provide a stringent setting in
which quantum interference must suppress radiative decay despite
photon--photon interactions. Here, we prove a no-go theorem showing that a
single nonlinear mode linearly coupled to a noninteracting bosonic continuum
cannot support an interaction-active two-photon bound state in the
continuum, provided the local spectral density vanishes only at isolated
energies. Apart from this regularity condition, the result is independent of
the continuum dispersion and the frequency-dependent coupling. Although such
an exact bound state is forbidden, long-lived two-photon quasi-bound states
remain possible. For a giant Kerr cavity nonlocally coupled to a waveguide,
we derive the asymptotic scaling of the two-photon decay rate with the
coupling-point separation from weak nonlinearity to the hard-core limit. We
further identify a regime in which the two-photon resonance is substantially
longer lived than the single-photon excitation. Our work establishes the
limits of exact multiphoton confinement while opening a route to engineering
long-lived interacting few-photon states.
\end{abstract}

\maketitle

\textit{Introduction---} Can destructive interference protect an interacting
photon pair even when its energy lies inside an open two-photon continuum?
This question extends the concept of a single-excitation bound state in the
continuum (BIC), a normalizable eigenstate degenerate with propagating modes
but decoupled from them by symmetry or interference. BICs have been widely
studied in photonic, acoustic, electronic, and matter-wave systems \cite%
{vonNeumann1929,Hsu2016,Sadreev2021,Kang2023}. Waveguide quantum
electrodynamics provides a particularly simple setting in which such states
arise from interference between radiative pathways \cite%
{Roy2017,Gu2017,Sheremet2023,Tufarelli2013,GonzalezBallestero2013,
Facchi2016,Calajo2019}. Giant emitters generalize this mechanism through
spatially separated coupling points, producing frequency-dependent decay,
time-delayed feedback, and dark states \cite%
{Kockum2014,Gustafsson2014,Guo2017,Kockum2018,Kannan2020,
Zanner2022,Roccati2024}. These interference effects have been used to
engineer and control single-photon BICs \cite%
{Wang2021,XuGuo2024,Zhang2024Phase,Guo2026,Chang2026Control}. In this
single-excitation sector the problem is linear, and the BIC condition
follows directly from cancellation of the coupling to resonant continuum
modes.

The two-photon problem is qualitatively different. A nonlinearity correlates
the photons and can bind them into composite few-photon states, as found in
single-emitter scattering, boson-impurity models, finite-bandwidth
waveguides, chiral waveguide QED, and strongly interacting Rydberg media
\cite%
{Zheng2010,Shi2016,Calajo2016,Mahmoodian2020,Firstenberg2013,Bienias2014,Liang2018}%
. Embedding such an interacting state inside an open continuum is
considerably more restrictive. Exact two-particle BICs have nevertheless
been found in systems with additional structure: retardation between distant
emitters can produce a two-photon BIC \cite{AlvarezGiron2024}, while
giant-atom coupling can support doublon BICs or bound states embedded in a
doublon continuum \cite{Rieck2025,Zhang2026}. Such constructions involve
multiple emitters, interactions distributed through the photonic medium, or
both. These results raise a more basic question: can a single local
nonlinear channel alone support an exact interaction-active two-photon BIC?

\begin{figure}[tbp]
\centering
\includegraphics[width=\linewidth]{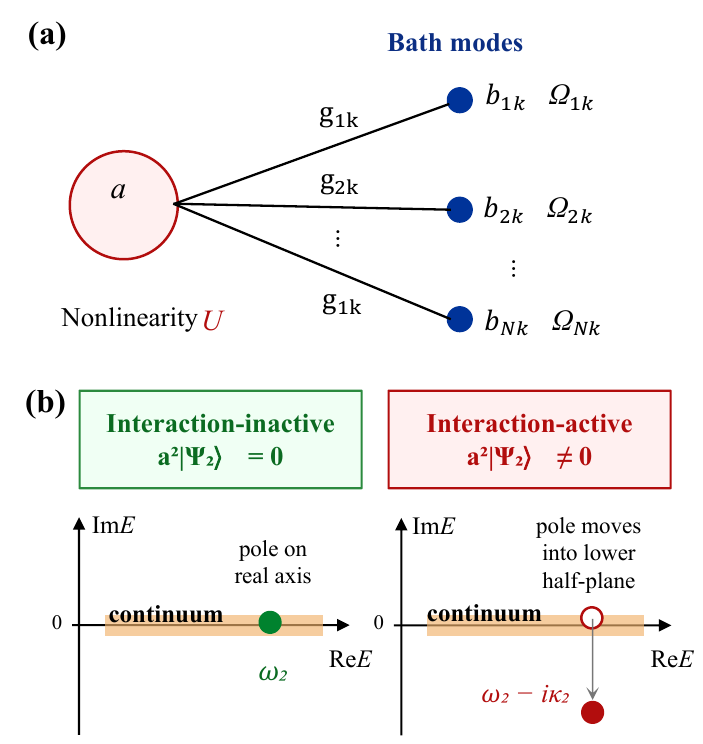}
\caption{Two-photon bound states in the continuum (BICs) for a single
nonlinear mode coupled to a bosonic continuum. (a) Schematic of a single
mode $a$ with nonlinearity $U$ linearly coupled to bosonic continuum modes $%
b_{jk}$ with dispersions $\Omega _{jk}$ and couplings $g_{jk}$. (b)
Classification of two-photon BICs. Interaction-active two-photon BICs are
excluded by the no-go theorem, whereas interaction-inactive two-photon BICs
may still exist.}
\label{fig1}
\end{figure}

We address this question in the minimal setting of a single nonlinear mode $%
a $ coupled linearly to a bosonic continuum with arbitrary dispersion and
frequency-dependent coupling [Fig.~\ref{fig1}(a)]. In this setting, the same
local two-photon amplitude that activates the interaction also couples the
state to the two-particle continuum. In few-photon scattering theory,
interaction-dependent two-particle poles are governed by the local pair
propagator and the corresponding $T$ matrix \cite%
{ShenFan2007,ShiChangCirac2015,Shi2016}. Using this structure, we prove a
no-go theorem: provided the local one-photon spectral density vanishes only
at isolated energies, a two-photon BIC with $a^{2}|\Psi_{2}\rangle\neq0$
cannot exist [Fig.~\ref{fig1}(b)]. By contrast, exact two-photon BICs may
survive when the interaction is inactive, $a^{2}|\Psi_{2}\rangle=0$ \cite%
{Crespi2015}.

The absence of an exact interaction-active BIC does not preclude a
long-lived two-excitation state. In contrast to the long-lived
two-excitation states found in emitter arrays as subradiant photon-pair
resonances, bound dimers, quasiflat-band states, and related in-band
resonances \cite{Ke2019,Zhang2020,Poddubny2020,Ke2020,Tecer2024}, we show
that a two-photon quasi-BIC can be supported by a single nonlinear mode
coupled to a linear continuum. Using a giant Kerr cavity nonlocally coupled
to a one-dimensional waveguide \cite{Chang2025NonMarkovian}, we derive its
asymptotic linewidth analytically, uncovering distinct scaling regimes with
the coupling-point separation from weak nonlinearity to the hard-core limit,
and verify these results from the exact interacting two-particle pole.

Finally, we show that frequency-selective interference can reverse the
lifetime hierarchy between the one- and two-particle resonances. The
nonlinearity shifts successive transitions to different frequencies,
allowing the one-photon transition to remain radiative while the dominant
two-photon decay channel lies close to an interference zero, analogous to
frequency-selective transition-rate engineering in multilevel emitters \cite%
{Kockum2014,Mirhosseini2018,Vadiraj2021}. Here, the linewidth of the
interacting two-photon quasi-BIC can be supressed, leading to a lifetime
greatly exceeding the single-photon lifetime. As a result, the linewidth of
the interacting two-photon quasi-BIC can be strongly suppressed, yielding a
lifetime that greatly exceeds the single-photon lifetime. This provides a
basis for engineering long-lived interacting few-photon states and for
identifying the minimal ingredients required for exact multiphoton
confinement in open quantum systems.

\textit{No-go theorem---} We consider a single nonlinear bosonic mode $a$
coupled linearly to $N$ bosonic continuum channels,
\begin{eqnarray}
H &=&\omega _{0}a^{\dagger }a+\frac{U}{2}a^{\dagger
2}a^{2}+\sum_{j=1}^{N}\int \mathrm{d}k\,\Omega _{jk}b_{jk}^{\dagger }b_{jk}
\notag \\
&&+\sum_{j=1}^{N}\int \mathrm{d}k\,\left( g_{jk}ab_{jk}^{\dagger }+\mathrm{%
H.c.}\right) .
\end{eqnarray}%
Here, $\omega _{0}$ is the frequency of the mode $a$, while $\Omega _{jk}$
and $g_{jk}$ denote the dispersion and coupling strength of the $j$th
continuum, respectively.

We first consider the single-excitation sector, for which the interaction $U$
is irrelevant. The Green's function of mode $a$ is $G_{a}(z)=[z-\omega
_{0}-\Sigma (z)]^{-1}$, with $\Sigma (z)=\sum_{j}\int \mathrm{d}%
k\,|g_{jk}|^{2}/(z-\Omega _{jk})$. The retarded Green's function is obtained
by taking $z=\omega +i0^{+}$. Its spectral representation can be written as
\begin{equation}
G_{a}(z)=\sum_{B}\frac{Z_{B}}{z-\omega _{B}}+\int_{\mathcal{C}}\mathrm{d}%
\omega \,\frac{\rho _{c}(\omega )}{z-\omega },  \label{1}
\end{equation}%
where the real poles $\omega _{B}$ satisfy $\omega _{B}-\omega _{0}-\Sigma
(\omega _{B})=0$ and have residues $Z_{B}=[1-\Sigma ^{\prime }(\omega
_{B})]^{-1}$. The continuum spectral density is $\rho _{c}(\omega )=-(1/\pi
) $Im$G_{a}(\omega +i0^{+})$, and $\mathcal{C}$ denotes the continuum
spectrum associated with the branch cuts of $G_{a}$. The pole and continuum
contributions correspond, respectively, to normalizable single-particle
bound states and scattering states. Their spectral weights are non-negative
and satisfy $\sum_{B}Z_{B}+\int_{\mathcal{C}}\mathrm{d}\omega \,\rho
_{c}(\omega )=1$.

An interaction-active two-photon BIC requires $U^{-1}-\Pi (z)=0$ to have a
real solution \cite{Shi2016,SupplementalMaterial}, where the pair bubble $%
\Pi (z)$ is the convolution of two single-particle Green's functions,
\begin{equation}
\Pi \left( z\right) =i\int G_{a}\left( \omega \right) G_{a}\left( z-\omega
\right) \frac{d\omega }{2\pi }.
\end{equation}%
Using Eq.~(\ref{1}), for $z\neq \omega _{B}+\omega _{B^{\prime }}$ we obtain
\begin{equation}
\text{Im}\Pi (z)=-2\pi \sum_{B}Z_{B}\rho _{c}(z-\omega _{B})-\pi \int_{%
\mathcal{C}}\rho _{c}(\omega )\rho _{c}(z-\omega )d\omega .
\end{equation}%
Since $Z_{B}\geq 0$ and $\rho _{c}(\omega )\geq 0$, inside the two-particle
continuum, we have $\text{Im}\Pi (E)<0$ provided $\rho _{c}(\omega )$
vanishes only at isolated frequencies. In terms of the coupling spectra $%
J_{j}(\omega )=\int dk\,|g_{jk}|^{2}\delta (\omega -\Omega _{jk})$, this
amounts to requiring that the total spectrum $J(\omega
)=\sum_{j}J_{j}(\omega )$ has only isolated zeros in the band, or,
equivalently, that the different continuum channels have only isolated
common zeros. At $z=\omega _{B}+\omega _{B^{\prime }}$, the
discrete--discrete contribution instead produces a pole of $\Pi (z)$, so
that $U^{-1}-\Pi (E)=0$ cannot be satisfied for nonzero $U$. This
singularity corresponds to the noninteracting product of two single-particle
bound states rather than an interaction-induced two-photon BIC. Thus, $%
U^{-1}-\Pi (E)=0$ has no real solution in the two-particle continuum, and an
interaction-active two-photon BIC is excluded.

Interaction-inactive multiphoton BICs may nevertheless exist when the linear
system supports more than one single-photon BIC \cite{SupplementalMaterial}.
Let $B_{n}^{\dagger }|0\rangle $ denote a single-photon BIC with energy $%
E_{n}$, satisfying $J(E_{n})=0$. In the two-excitation sector, for a fixed
total energy $E$, a general state constructed from single-photon BICs can be
written as $\left\vert \Psi _{2}\right\rangle =\sum_{mn}C_{mn}B_{m}^{\dagger
}B_{n}^{\dagger }|0\rangle $, where the sum is restricted to pairs
satisfying $E_{m}+E_{n}=E$. Such a state remains an exact two-photon BIC in
the presence of the nonlinearity provided $a^{2}|\Psi _{2}\rangle =0$, so
that the interaction is inactive in the state. If there are $N_{E}$ linearly
independent BIC-pair states at total energy $E$ and at least one
participating pair has nonzero $a$-mode weight, the condition $a^{2}|\Psi
_{2}\rangle =0$ imposes one independent linear constraint. The number of
interaction-inactive two-photon BICs at energy $E$ is therefore $N_{E}-1$.

\textit{Two-photon quasi-BICs---} In contrast to emitter arrays, where
long-lived two-excitation states can arise from collective effects \cite%
{Zhang2020,Poddubny2020,Ke2019,Tecer2024}, we show that a single nonlinear
mode can also support a long-lived two-photon quasi-BIC. As a concrete
example, we consider a giant Kerr cavity coupled nonlocally to a
one-dimensional waveguide at $M$ equally spaced coupling points \cite%
{Kockum2014,ZhaoWang2020,Guo2020BIC,Guo2020Oscillating,Gu2024,
Chang2025NonMarkovian}, and study how the decay rate $\kappa _{2}$ of the
two-photon quasi-BIC scales with $M$ and the separation between the coupling
points.

Let the separation between neighboring coupling points be $d$. The left- and
right-moving waveguide modes have dispersions $\Omega _{1k}=k_{0}-k$ and $%
\Omega _{2k}=k_{0}+k$, respectively, where the central frequency $k_{0}$ is
chosen to be $\omega _{0}$. The corresponding couplings are $%
g_{1k}=g_{0}\sum_{m=1}^{M}e^{im(k_{0}-k)d}$ and $g_{2k}=g_{0}%
\sum_{m=1}^{M}e^{-im(k_{0}+k)d}$, with $g_{0}$ the coupling strength at each
coupling point. The self-energy can then be written as $\Sigma (z)=\Delta
(z)-i\Gamma (z)$ \cite{Chang2025NonMarkovian}, with
\begin{equation}
\Delta \left( z\right) =\gamma \frac{M\sin zd-\sin Mzd}{2\sin ^{2}\left(
zd/2\right) }
\end{equation}%
and
\begin{equation}
\Gamma \left( z\right) =\gamma \frac{\sin ^{2}\left( Mzd/2\right) }{\sin
^{2}\left( zd/2\right) },
\end{equation}%
where $\gamma =2\pi g_{0}^{2}$. A single-photon BIC requires \cite%
{Kockum2014,Guo2020BIC,Guo2020Oscillating} $\omega _{B}-\omega _{0}-\Sigma
(\omega _{B})=0$. This gives the BIC energies
\begin{equation}
\omega _{Bq}d=\frac{2\pi q}{M}+2\pi n,\qquad q=1,\ldots ,M-1,  \label{2}
\end{equation}%
together with the corresponding condition on the cavity frequency,
\begin{equation}
\omega _{0}=\omega _{Bq}-\gamma M\cot \frac{q\pi }{M}.  \label{3}
\end{equation}

\begin{figure}[tbp]
\centering
\includegraphics[width=\linewidth]{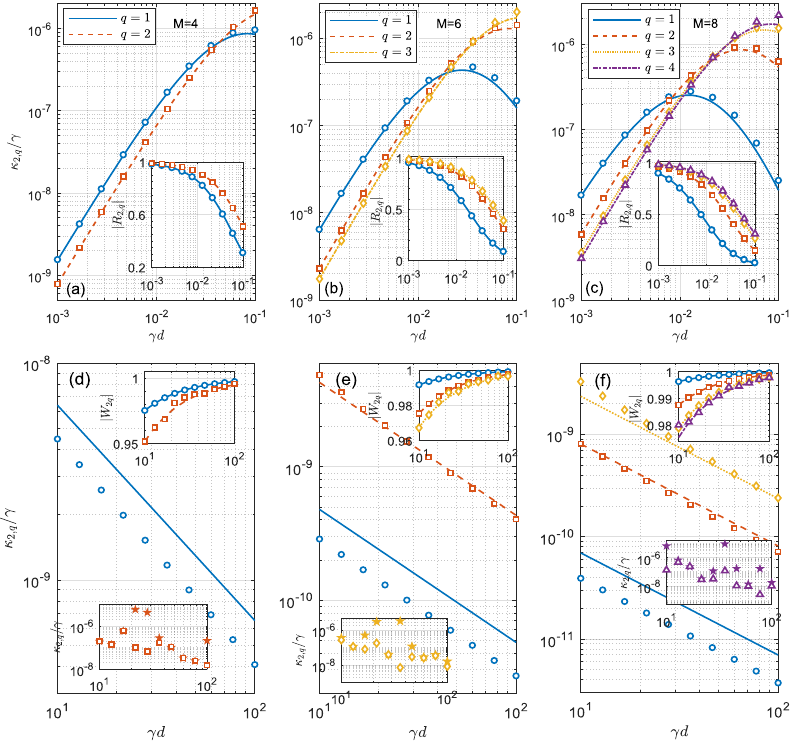}
\caption{Two-photon quasi-BICs in the weak-nonlinearity limit. (a)--(c)
Decay rate $\protect\kappa _{2}$ in the small-$d$ regime. Lines show the
analytical results, while circles, squares, diamonds, and triangles denote
the numerical results for $q=1,2,3,4$, respectively. The same symbols are
used in the insets. (d)--(f) Decay rate $\protect\kappa _{2}$ in the large-$d
$ regime. The notation is the same as in (a)--(c), except for the $q=M/2$
results shown in the lower insets, where pentagrams denote the analytical
results.}
\label{fig2}
\end{figure}

The two-photon quasi-BICs are determined by the poles satisfying $U^{-1}-\Pi
(\omega _{2}-i\kappa _{2})=0$, where $\omega _{2}$ and $\kappa _{2}$ are the
corresponding energy and decay rate, respectively \cite{SupplementalMaterial}%
. There are two classes of solutions. One is continuously connected to
combinations of single-photon quasi-BICs, while the other is induced by the
nonlinearity and satisfies $\lim_{U\rightarrow 0}\kappa _{2}=0$. We focus on
the latter, since it directly characterizes the effect of the interaction
between two BIC photons.\textquotedblright .

In the weak-nonlinearity limit, one finds $\kappa_{2}(q)=\kappa_{2}(M-q)$.
For $M\gamma d\ll s_{q}^{2}$,
\begin{equation}
\kappa _{2}^{\text{small }d}\approx \frac{U^{2}}{\gamma }\frac{\left(
M\gamma d\right) ^{2}}{2s_{q}^{2}}Z_{B}^{5},
\end{equation}
where $s_{q}=\sin(q\pi/M)$ and $Z_{Bq}=(1+M\gamma d/s_{q}^{2})^{-1}$. Thus, $%
\kappa_{2}^{\text{small }d}$ increases with $d$, as shown in Figs.~\ref{fig2}%
(a)--(c) for $U=0.01\gamma$. This behavior can be understood from the
phase-matching condition. The single-photon BIC is protected by the
interference condition in Eq.~(\ref{2}), whereas the interaction shifts the
two-photon phase away from this condition by an amount $UdZ_{B}^{2}$. In the
small-$d$ limit, this shift reduces to approximately $Ud$, and therefore
vanishes as $d$ decreases.

We characterize the quasi-BIC by the magnitude $|R_{2q}|$ of the
two-cavity-photon Green-function residue and by the two-waveguide-photon
weight $|W_{2q}|$ inside the coupling region (concrete definitions are given
in the Supplemental Material \cite{SupplementalMaterial}). The cavity
component $R_{2q}$, shown in the insets, decreases with increasing $d$, as
in the single-photon BIC. Correspondingly, the two-photon waveguide
component $W_{2q}$ confined within $[0,(M-1)d]$ increases, as shown in the
upper insets of Figs.~\ref{fig2}(d)--(e). In contrast to the small-$d$
limit, in large-$d$ limit, the interaction-induced phase shift scales
instead as $Us_{q}^{4}/(M^{2}\gamma ^{2}d)$, so that the decay decreases
with increasing $d$. For $M\gamma d\gg s_{q}^{2}$,
\begin{equation}
\kappa _{2}^{\text{large }d}\approx U^{2}Z_{B}^{2}\left[ \frac{\gamma
M^{2}d^{2}}{2s_{q}^{2}}Z_{B}^{3}+\frac{d}{6M}+\xi \delta _{q,M/2}\right] ,
\end{equation}%
where the third term $\xi =\frac{d}{2M}\sum_{p\neq M/2}^{M-1}\csc ^{2}\left(
\frac{p\pi }{M}-\frac{\pi }{2}-\frac{1}{2}M\gamma d\cot \frac{p\pi }{M}%
\right) $ is present only for $q=M/2$. The corresponding $\kappa _{2}$ for $%
q=M/2$ are shown in the lower insets of Figs.~\ref{fig2}(d)--(e). In this
case, the analytical expression exhibits oscillations and deviates from the
numerical result at a few isolated values of $d$, where $U=0.01\gamma $ is
not sufficiently small for the perturbative expansion to remain accurate.

\begin{figure}[tbp]
\centering
\includegraphics[width=\linewidth]{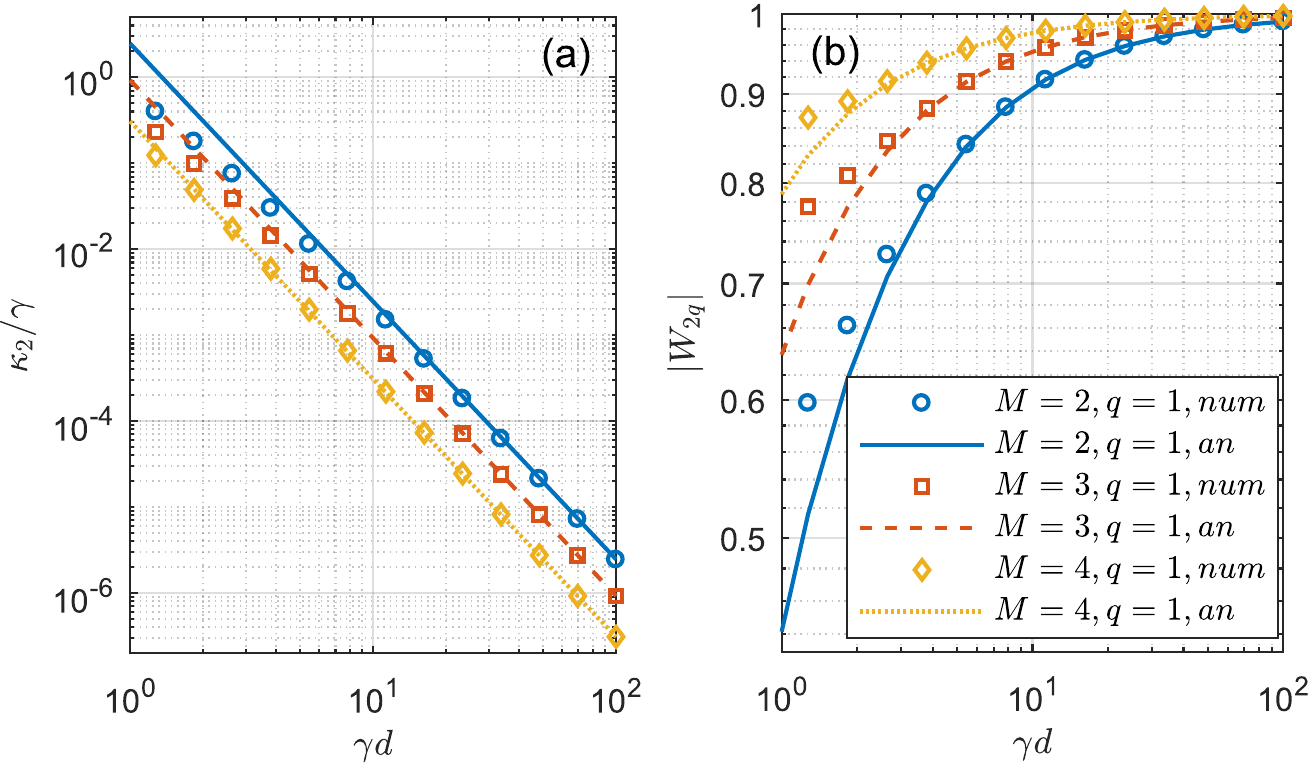}
\caption{Two-photon quasi-BICs in hard-core limit. (a) Decay rate $\protect%
\kappa_{2}$ of the two-photon quasi-BIC in the hard-core limit $%
U\rightarrow+\infty$. Markers show the numerical results obtained from the
pole condition $\Pi(z)=0$, while solid lines denote the analytical results.
The symmetry $\protect\kappa_{2}(q)=\protect\kappa_{2}(M-q)$ has been used.
(b) Corresponding two-waveguide-photon weight $W_{2q}$ within the
interaction region.}
\label{fig3}
\end{figure}

We have shown that a weak nonlinearity generates two-photon quasi-BICs with
a decay rate scaling as $U^{2}$. Long-lived two-photon quasi-BICs also
persist in the opposite, strongly interacting regime. As shown in Fig.~\ref%
{fig3}(a), in the hard-core limit $U\rightarrow +\infty $, the decay rate $%
\kappa _{2}$ decreases with increasing $d$. For $q\neq M/2$, as well as for
the special case $M=2$, $q=1$, its large-$d$ asymptotic form is
\begin{equation}
\kappa _{2}^{\text{hard-core}}\approx \gamma \frac{\pi ^{2}s_{q}^{4}}{%
2M\left( \gamma d\right) ^{3}}.
\end{equation}%
Thus, even in the hard-core limit, $\kappa _{2}\ll \gamma $ can be achieved
by increasing the separation between the coupling points. This suppression
is associated with the decreasing $a$-mode weight of the underlying
single-photon BIC as $d$ increases. For fixed $q$, increasing $M$ further
reduces $\kappa _{2}$, consistent with the smaller single-photon residue $%
Z_{B}$. Figure~\ref{fig3}(b) shows the corresponding two-photon waveguide
weight $W_{2q}$, which increases with both $d$ and $M$. For even $M>2$, the
symmetry branch $q=M/2$ contains additional degenerate pair channels and is
discussed separately in the Supplemental Material.

Thus, long-lived two-photon quasi-BICs persist from weak nonlinearity to the
hard-core limit. Their decay shows distinct small- and large-$d$ behaviors,
but can become parametrically smaller than the bare decay rate at large
separation, demonstrating that a single nonlinear mode is sufficient to
support a strongly confined two-photon resonance.

\textit{Away from the single-photon-BIC condition---} We have shown that, at
the single-photon BIC condition, a long-lived two-excitation state can
persist even in the strongly interacting regime. We now move away from the
condition in Eq.~(\ref{3}) and show that the nonlinearity allows the one-
and two-photon decay rates to be tuned differently. In particular, a regime
with $\kappa _{1}\gg \kappa _{2}$ can be reached, reversing the hierarchy at
the single-photon BIC point \cite{SupplementalMaterial}.

We introduce a detuning $\delta_{0}$ from the single-photon BIC condition,
\begin{equation}
\delta _{0}=\omega _{0}-\omega _{Bq}+M\gamma \cot \frac{q\pi }{M}.
\end{equation}
The corresponding single-photon quasi-BIC is determined by $%
\omega_{1}-i\kappa_{1}-\omega_{0} -\Sigma(\omega_{1}-i\kappa_{1})=0$, where $%
\omega_{1}$ and $\kappa_{1}$ are its energy and decay rate. The origin of
the different decay rates can be understood from the successive one-to-zero
and two-to-one transitions, whose bare transition frequencies are $\omega_0$
and $\omega_0+U$, respectively. To realize $\kappa_{1}\gg\kappa_{2}$, we
choose $\delta_{0}\neq0$, so that the one-to-zero transition is away from
the single-photon BIC condition, while tuning $U+\delta_{0}\simeq0$ so that
the two-to-one transition is close to the corresponding BIC condition.

A smaller $d$ can further increase the ratio $\kappa _{1}/\kappa _{2}$,
because it makes the one-to-zero and two-to-one decay channels increasingly
dominated by their respective transition frequencies. But the limit $%
d\rightarrow 0$ cannot distinguish the two decay channels. A finite
propagation phase is required both to make the one-to-zero transition bright
and to keep the two-to-one transition dark. Thus, the regime $\kappa _{1}\gg
\kappa _{2}$ requires a finite, although possibly small, separation between
the coupling points.

\begin{figure}[tbp]
\centering
\includegraphics[width=\linewidth]{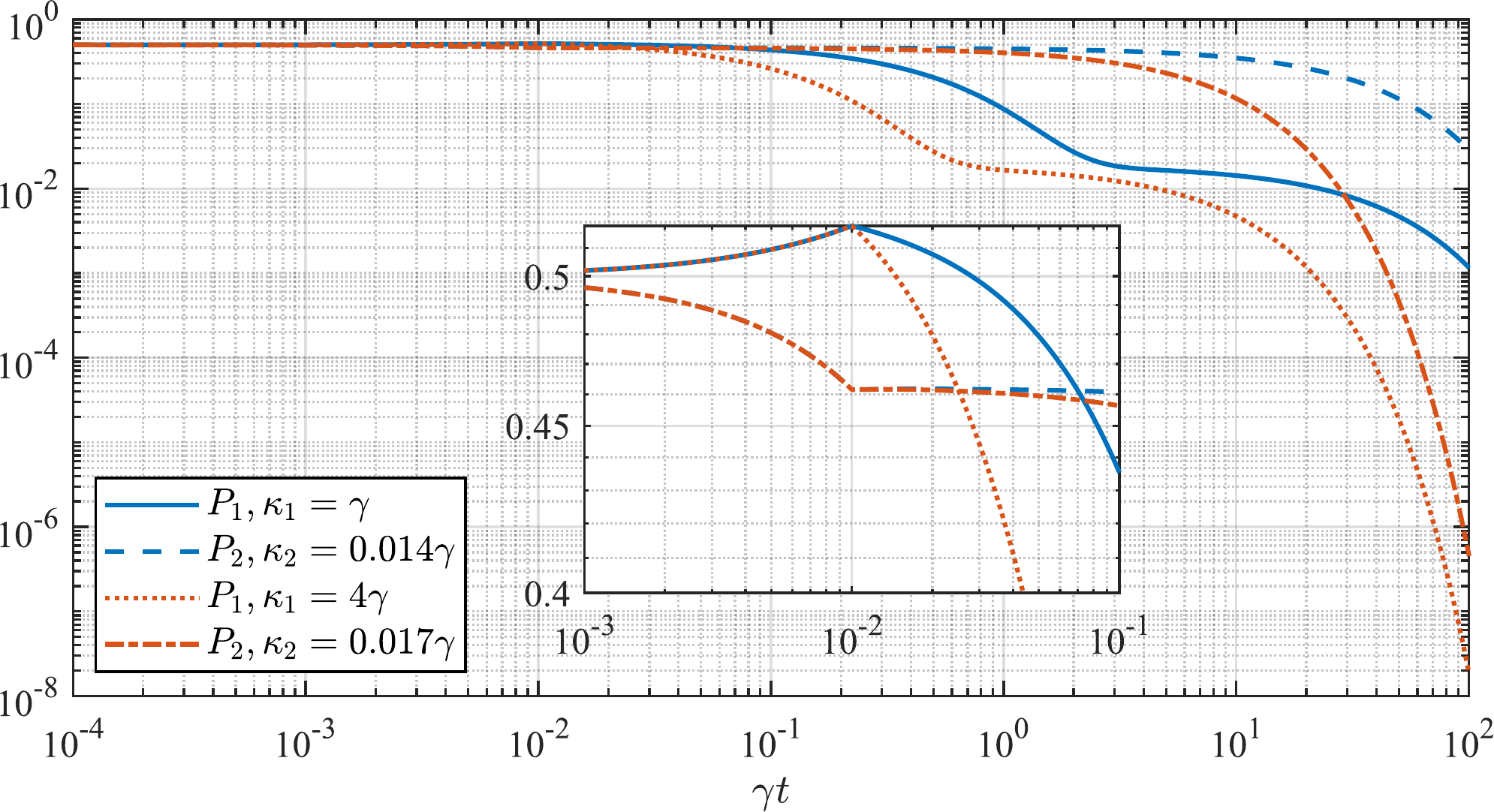}
\caption{Decay of the single- and two-photon populations $P_{1}$ and $P_{2}$
of the $a$ mode. Solid and dashed lines correspond to $U=-\protect\delta %
_{0}=107\protect\gamma $, while dotted and dash-dotted lines correspond to $%
U=-\protect\delta _{0}=287\protect\gamma $.}
\label{fig4}
\end{figure}

In Fig.~\ref{fig4}, we show the single- and two-photon populations $P_{1}$
and $P_{2}$ of the $a$ mode for the initial state $\left( a^{\dagger }/\sqrt{%
2}+a^{\dagger 2}/2\right) |0\rangle $, with $M=2$, $\gamma d=0.01$, and $%
U=-\delta _{0}=107\gamma $ or $287\gamma $. For these parameters, $\kappa
_{1}\sim \gamma $, whereas $\kappa _{2}\sim 0.01\gamma $. At short times, $%
P_{1}$ increases slightly because of the weak decay of the two-photon
component into one $a$-mode photon and one waveguide photon. Around $\gamma
t=\gamma d=0.01$, corresponding to the propagation time between neighboring
coupling points, interference is established. Thereafter, the contribution
originating from the initial one-photon sector decays at the rate $2\kappa
_{1}$, causing $P_{1}$ to decrease rapidly, whereas $P_{2}$ decays much more
slowly at the rate $2\kappa _{2}$. The single-photon component can therefore
be efficiently removed while leaving the two-photon component nearly
unchanged over the same time interval, providing an efficient way for
excitation-selective confinement.

\textit{Conclusion and outlook---} We have established a fundamental
constraint on two-photon BICs in the minimal setting of a single nonlinear
bosonic mode coupled linearly to a noninteracting continuum. Using the exact
two-particle Green function, we proved that, provided the local one-photon
spectral density vanishes only at isolated energies, an interaction-active
two-photon BIC cannot exist, independently of the bath dispersion and the
frequency dependence of the coupling. Exact two-photon BICs may nevertheless
survive when the interaction annihilates the state, but they do not
represent interaction-induced confinement.

The absence of an exact interaction-active BIC does not preclude a
parametrically long-lived resonance. For a giant Kerr cavity coupled
nonlocally to a waveguide, we identified two-photon quasi-BICs and derived
the asymptotic scaling of their linewidth with the coupling-point separation
from weak nonlinearity to the hard-core limit. We further showed that
frequency-selective interference can reverse the lifetime hierarchy between
the one- and two-photon resonances, yielding regimes with $%
\kappa_{2}\ll\kappa_{1}$. Thus, even with only one nonlinear degree of
freedom, two excitations can remain confined substantially longer than one.

The no-go theorem established here also points to a route beyond the
single-mode setting. With several nonlinear degrees of freedom, the local
pair propagator becomes matrix valued, allowing interference between
distinct interaction channels that is absent in the rank-one problem
considered here. A natural next step is to establish general criteria for
such multichannel systems and to extend the corresponding constraints to
higher excitation sectors. In addition, the reversed lifetime hierarchy
demonstrated here suggests excitation-selective confinement, in which a
chosen multiphoton sector is long lived while other excitation sectors
remain radiative.

\textit{Acknowledgments.---} This work was supported by the National Natural
Science Foundation of China under Grant No.~12575008.

\bibliographystyle{unsrtnat}
\bibliography{ref}

\begin{thebibliography}{49}
\providecommand{\natexlab}[1]{#1}
\providecommand{\url}[1]{\texttt{#1}}
\expandafter\ifx\csname urlstyle\endcsname\relax
  \providecommand{\doi}[1]{doi: #1}\else
  \providecommand{\doi}{doi: \begingroup \urlstyle{rm}\Url}\fi

\bibitem[von Neumann and Wigner(1929)]{vonNeumann1929}
John von Neumann and Eugene~P. Wigner.
\newblock {\"U}ber merkw{\"u}rdige diskrete eigenwerte.
\newblock \emph{Physikalische Zeitschrift}, 30:\penalty0 465--467, 1929.

\bibitem[Hsu et~al.(2016)Hsu, Zhen, Stone, Joannopoulos, and
  Solja{\v{c}}i{\'c}]{Hsu2016}
Chia~Wei Hsu, Bo~Zhen, A.~Douglas Stone, John~D. Joannopoulos, and Marin
  Solja{\v{c}}i{\'c}.
\newblock Bound states in the continuum.
\newblock \emph{Nature Reviews Materials}, 1:\penalty0 16048, 2016.
\newblock \doi{10.1038/natrevmats.2016.48}.

\bibitem[Sadreev(2021)]{Sadreev2021}
Almas~F. Sadreev.
\newblock Interference traps waves in an open system: Bound states in the
  continuum.
\newblock \emph{Reports on Progress in Physics}, 84\penalty0 (5):\penalty0
  055901, 2021.
\newblock \doi{10.1088/1361-6633/abefb9}.

\bibitem[Kang et~al.(2023)Kang, Liu, Chan, and Xiao]{Kang2023}
Meng Kang, Tao Liu, C.~T. Chan, and Meng Xiao.
\newblock Applications of bound states in the continuum in photonics.
\newblock \emph{Nature Reviews Physics}, 5\penalty0 (11):\penalty0 659--678,
  2023.
\newblock \doi{10.1038/s42254-023-00642-8}.

\bibitem[Roy et~al.(2017)Roy, Wilson, and Firstenberg]{Roy2017}
Dibyendu Roy, C.~M. Wilson, and Ofer Firstenberg.
\newblock Colloquium: Strongly interacting photons in one-dimensional
  continuum.
\newblock \emph{Reviews of Modern Physics}, 89\penalty0 (2):\penalty0 021001,
  2017.
\newblock \doi{10.1103/RevModPhys.89.021001}.

\bibitem[Gu et~al.(2017)Gu, Kockum, Miranowicz, Liu, and Nori]{Gu2017}
Xiu Gu, Anton~Frisk Kockum, Adam Miranowicz, Yu-xi Liu, and Franco Nori.
\newblock Microwave photonics with superconducting quantum circuits.
\newblock \emph{Physics Reports}, 718--719:\penalty0 1--102, 2017.
\newblock \doi{10.1016/j.physrep.2017.10.002}.

\bibitem[Sheremet et~al.(2023)Sheremet, Petrov, Iorsh, Poshakinskiy, and
  Poddubny]{Sheremet2023}
Alexandra~S. Sheremet, Mihail~I. Petrov, Ivan~V. Iorsh, Alexander~V.
  Poshakinskiy, and Alexander~N. Poddubny.
\newblock Waveguide quantum electrodynamics: Collective radiance and
  photon-photon correlations.
\newblock \emph{Reviews of Modern Physics}, 95\penalty0 (1):\penalty0 015002,
  2023.
\newblock \doi{10.1103/RevModPhys.95.015002}.

\bibitem[Tufarelli et~al.(2013)Tufarelli, Ciccarello, and Kim]{Tufarelli2013}
Tommaso Tufarelli, Francesco Ciccarello, and M.~S. Kim.
\newblock Dynamics of spontaneous emission in a single-end photonic waveguide.
\newblock \emph{Physical Review A}, 87\penalty0 (1):\penalty0 013820, 2013.
\newblock \doi{10.1103/PhysRevA.87.013820}.

\bibitem[Gonz{\'a}lez-Ballestero et~al.(2013)Gonz{\'a}lez-Ballestero,
  Garc{\'i}a-Vidal, and Moreno]{GonzalezBallestero2013}
Carlos Gonz{\'a}lez-Ballestero, Francisco~J. Garc{\'i}a-Vidal, and Esteban
  Moreno.
\newblock Non-{M}arkovian effects in waveguide-mediated entanglement.
\newblock \emph{New Journal of Physics}, 15\penalty0 (7):\penalty0 073015,
  2013.
\newblock \doi{10.1088/1367-2630/15/7/073015}.

\bibitem[Facchi et~al.(2016)Facchi, Kim, Pascazio, Pepe, Pomarico, and
  Tufarelli]{Facchi2016}
Paolo Facchi, M.~S. Kim, Saverio Pascazio, Francesco~V. Pepe, Domenico
  Pomarico, and Tommaso Tufarelli.
\newblock Bound states and entanglement generation in waveguide quantum
  electrodynamics.
\newblock \emph{Physical Review A}, 94\penalty0 (4):\penalty0 043839, 2016.
\newblock \doi{10.1103/PhysRevA.94.043839}.

\bibitem[Calaj{\'o} et~al.(2019)Calaj{\'o}, Fang, Baranger, and
  Ciccarello]{Calajo2019}
Giuseppe Calaj{\'o}, Yao-Lung~L. Fang, Harold~U. Baranger, and Francesco
  Ciccarello.
\newblock Exciting a bound state in the continuum through multiphoton
  scattering plus delayed quantum feedback.
\newblock \emph{Physical Review Letters}, 122\penalty0 (7):\penalty0 073601,
  2019.
\newblock \doi{10.1103/PhysRevLett.122.073601}.

\bibitem[Kockum et~al.(2014)Kockum, Delsing, and Johansson]{Kockum2014}
Anton~Frisk Kockum, Per Delsing, and G{\"o}ran Johansson.
\newblock Designing frequency-dependent relaxation rates and lamb shifts for a
  giant artificial atom.
\newblock \emph{Physical Review A}, 90\penalty0 (1):\penalty0 013837, 2014.
\newblock \doi{10.1103/PhysRevA.90.013837}.

\bibitem[Gustafsson et~al.(2014)Gustafsson, Aref, Kockum, Ekstr{\"o}m,
  Johansson, and Delsing]{Gustafsson2014}
Martin~V. Gustafsson, Thomas Aref, Anton~Frisk Kockum, Maria~K. Ekstr{\"o}m,
  G{\"o}ran Johansson, and Per Delsing.
\newblock Propagating phonons coupled to an artificial atom.
\newblock \emph{Science}, 346\penalty0 (6206):\penalty0 207--211, 2014.
\newblock \doi{10.1126/science.1257219}.

\bibitem[Guo et~al.(2017)Guo, Grimsmo, Kockum, Pletyukhov, and
  Johansson]{Guo2017}
Lingzhen Guo, Arne~L. Grimsmo, Anton~Frisk Kockum, Mikhail Pletyukhov, and
  G{\"o}ran Johansson.
\newblock Giant acoustic atom: A single quantum system with a deterministic
  time delay.
\newblock \emph{Physical Review A}, 95\penalty0 (5):\penalty0 053821, 2017.
\newblock \doi{10.1103/PhysRevA.95.053821}.

\bibitem[Kockum et~al.(2018)Kockum, Johansson, and Nori]{Kockum2018}
Anton~Frisk Kockum, G{\"o}ran Johansson, and Franco Nori.
\newblock Decoherence-free interaction between giant atoms in waveguide quantum
  electrodynamics.
\newblock \emph{Physical Review Letters}, 120\penalty0 (14):\penalty0 140404,
  2018.
\newblock \doi{10.1103/PhysRevLett.120.140404}.

\bibitem[Kannan et~al.(2020)Kannan, Ruckriegel, Campbell, Kockum,
  Braum{\"u}ller, Kim, Kjaergaard, Krantz, Melville, Niedzielski,
  Veps{\"a}l{\"a}inen, Winik, Yoder, Nori, Orlando, Gustavsson, and
  Oliver]{Kannan2020}
Bharath Kannan, M.~J. Ruckriegel, D.~L. Campbell, Anton~Frisk Kockum, Jochen
  Braum{\"u}ller, D.~K. Kim, Morten Kjaergaard, Philip Krantz, Alexander
  Melville, Bethany~M. Niedzielski, Antti Veps{\"a}l{\"a}inen, Roni Winik,
  J.~L. Yoder, Franco Nori, T.~P. Orlando, Simon Gustavsson, and William~D.
  Oliver.
\newblock Waveguide quantum electrodynamics with superconducting artificial
  giant atoms.
\newblock \emph{Nature}, 583:\penalty0 775--779, 2020.
\newblock \doi{10.1038/s41586-020-2529-9}.

\bibitem[Zanner et~al.(2022)Zanner, Orell, Schneider, Albert, Oleschko, Juan,
  Silveri, and Kirchmair]{Zanner2022}
Maximilian Zanner, Tuure Orell, Christian M.~F. Schneider, Romain Albert,
  Stefan Oleschko, Mathieu~L. Juan, Matti Silveri, and Gerhard Kirchmair.
\newblock Coherent control of a multi-qubit dark state in waveguide quantum
  electrodynamics.
\newblock \emph{Nature Physics}, 18:\penalty0 538--543, 2022.
\newblock \doi{10.1038/s41567-022-01527-w}.

\bibitem[Roccati and Cilluffo(2024)]{Roccati2024}
Federico Roccati and Dario Cilluffo.
\newblock Controlling markovianity with chiral giant atoms.
\newblock \emph{Physical Review Letters}, 133\penalty0 (6):\penalty0 063603,
  2024.
\newblock \doi{10.1103/PhysRevLett.133.063603}.

\bibitem[Wang et~al.(2021)Wang, Liu, Kockum, Li, and Nori]{Wang2021}
Xin Wang, Tao Liu, Anton~Frisk Kockum, Hong-Rong Li, and Franco Nori.
\newblock Tunable chiral bound states with giant atoms.
\newblock \emph{Physical Review Letters}, 126\penalty0 (4):\penalty0 043602,
  2021.
\newblock \doi{10.1103/PhysRevLett.126.043602}.

\bibitem[Xu and Guo(2024)]{XuGuo2024}
Luting Xu and Lingzhen Guo.
\newblock Catch and release of propagating bosonic field with non-{M}arkovian
  giant atom.
\newblock \emph{New Journal of Physics}, 26\penalty0 (1):\penalty0 013025,
  2024.
\newblock \doi{10.1088/1367-2630/ad18ed}.

\bibitem[Zhang et~al.(2025)Zhang, Zhu, and Wang]{Zhang2024Phase}
Xiaojun Zhang, Mingjie Zhu, and Zhihai Wang.
\newblock Phase-controlled bound states in giant atom waveguide {QED} setup.
\newblock \emph{Communications in Theoretical Physics}, 77\penalty0
  (11):\penalty0 115102, 2025.
\newblock \doi{10.1088/1572-9494/addb2a}.

\bibitem[Guo et~al.(2026)Guo, Zhang, Weng, Bin, Liu, Xing, L{\"u}, and
  Wang]{Guo2026}
Xiang Guo, Xiaojun Zhang, Mingzhu Weng, Qian Bin, Hao-di Liu, Hai-Jun Xing,
  Xin-You L{\"u}, and Zhihai Wang.
\newblock Bound state in the continuum and multiple atom state transfer
  applications in a waveguide qed setup.
\newblock \emph{Physical Review Letters}, 137:\penalty0 073601, Aug 2026.
\newblock \doi{10.1103/y1rg-p9hx}.
\newblock URL \url{https://link.aps.org/doi/10.1103/y1rg-p9hx}.
\newblock Accepted 10 July 2026.

\bibitem[Chang(2026)]{Chang2026Control}
Yue Chang.
\newblock Coherent control of an embedded bound state without a spectral gap.
\newblock \emph{arXiv preprint arXiv:2606.17685}, 2026.
\newblock \doi{10.48550/arXiv.2606.17685}.

\bibitem[Zheng et~al.(2010)Zheng, Gauthier, and Baranger]{Zheng2010}
Huaixiu Zheng, Daniel~J. Gauthier, and Harold~U. Baranger.
\newblock Waveguide {QED}: Many-body bound-state effects in coherent and
  fock-state scattering from a two-level system.
\newblock \emph{Physical Review A}, 82\penalty0 (6):\penalty0 063816, 2010.
\newblock \doi{10.1103/PhysRevA.82.063816}.

\bibitem[Shi et~al.(2016)Shi, Wu, Gonz{\'a}lez-Tudela, and Cirac]{Shi2016}
Tao Shi, Ying-Hai Wu, Alejandro Gonz{\'a}lez-Tudela, and J.~Ignacio Cirac.
\newblock Bound states in boson impurity models.
\newblock \emph{Physical Review X}, 6\penalty0 (2):\penalty0 021027, 2016.
\newblock \doi{10.1103/PhysRevX.6.021027}.

\bibitem[Calaj{\'o} et~al.(2016)Calaj{\'o}, Ciccarello, Chang, and
  Rabl]{Calajo2016}
Giuseppe Calaj{\'o}, Francesco Ciccarello, Darrick~E. Chang, and Peter Rabl.
\newblock Atom-field dressed states in slow-light waveguide {QED}.
\newblock \emph{Physical Review A}, 93\penalty0 (3):\penalty0 033833, 2016.
\newblock \doi{10.1103/PhysRevA.93.033833}.

\bibitem[Mahmoodian et~al.(2020)Mahmoodian, Calaj{\'o}, Chang, Hammerer, and
  S{\o}rensen]{Mahmoodian2020}
Sahand Mahmoodian, Giuseppe Calaj{\'o}, Darrick~E. Chang, Klemens Hammerer, and
  Anders~S. S{\o}rensen.
\newblock Dynamics of many-body photon bound states in chiral waveguide {QED}.
\newblock \emph{Physical Review X}, 10\penalty0 (3):\penalty0 031011, 2020.
\newblock \doi{10.1103/PhysRevX.10.031011}.

\bibitem[Firstenberg et~al.(2013)Firstenberg, Peyronel, Liang, Gorshkov, Lukin,
  and Vuleti{\'c}]{Firstenberg2013}
Ofer Firstenberg, Thibault Peyronel, Qi-Yu Liang, Alexey~V. Gorshkov,
  Mikhail~D. Lukin, and Vladan Vuleti{\'c}.
\newblock Attractive photons in a quantum nonlinear medium.
\newblock \emph{Nature}, 502:\penalty0 71--75, 2013.
\newblock \doi{10.1038/nature12512}.

\bibitem[Bienias et~al.(2014)Bienias, Choi, Firstenberg, Maghrebi, Gullans,
  Lukin, Gorshkov, and B{\"u}chler]{Bienias2014}
P.~Bienias, S.~Choi, O.~Firstenberg, M.~F. Maghrebi, M.~Gullans, M.~D. Lukin,
  A.~V. Gorshkov, and H.~P. B{\"u}chler.
\newblock Scattering resonances and bound states for strongly interacting
  {Rydberg} polaritons.
\newblock \emph{Physical Review A}, 90\penalty0 (5):\penalty0 053804, 2014.
\newblock \doi{10.1103/PhysRevA.90.053804}.

\bibitem[Liang et~al.(2018)Liang, Venkatramani, Cantu, Nicholson, Gullans,
  Gorshkov, Thompson, Chin, Lukin, and Vuleti{\'c}]{Liang2018}
Qi-Yu Liang, Aditya~V. Venkatramani, Sergio~H. Cantu, Travis~L. Nicholson,
  Michael~J. Gullans, Alexey~V. Gorshkov, Jeff~D. Thompson, Cheng Chin,
  Mikhail~D. Lukin, and Vladan Vuleti{\'c}.
\newblock Observation of three-photon bound states in a quantum nonlinear
  medium.
\newblock \emph{Science}, 359\penalty0 (6377):\penalty0 783--786, 2018.
\newblock \doi{10.1126/science.aao7293}.

\bibitem[Alvarez-Giron et~al.(2024)Alvarez-Giron, Solano, Sinha, and
  Barberis-Blostein]{AlvarezGiron2024}
W.~Alvarez-Giron, P.~Solano, K.~Sinha, and P.~Barberis-Blostein.
\newblock Delay-induced spontaneous dark-state generation from two distant
  excited atoms.
\newblock \emph{Physical Review Research}, 6\penalty0 (2):\penalty0 023213,
  2024.
\newblock \doi{10.1103/PhysRevResearch.6.023213}.

\bibitem[Rieck et~al.(2025)Rieck, Kockum, and Chen]{Rieck2025}
Walter Rieck, Anton~Frisk Kockum, and Guangze Chen.
\newblock Doublon bound states in the continuum through giant atoms.
\newblock \emph{arXiv preprint arXiv:2511.18212}, 2025.
\newblock \doi{10.48550/arXiv.2511.18212}.

\bibitem[Zhang et~al.(2026)Zhang, Guo, Zhang, Wang, Xing, and Wang]{Zhang2026}
Xiaojun Zhang, Xiang Guo, Yan Zhang, Xin Wang, Haijun Xing, and Zhihai Wang.
\newblock Quantum state preparation and transfer based on the bound state in
  the doublon continuum.
\newblock \emph{Physical Review A}, 113\penalty0 (5):\penalty0 053714, 2026.
\newblock \doi{10.1103/yhn3-vnyj}.

\bibitem[Shen and Fan(2007)]{ShenFan2007}
Jung-Tsung Shen and Shanhui Fan.
\newblock Strongly correlated two-photon transport in a one-dimensional
  waveguide coupled to a two-level system.
\newblock \emph{Physical Review Letters}, 98\penalty0 (15):\penalty0 153003,
  2007.
\newblock \doi{10.1103/PhysRevLett.98.153003}.

\bibitem[Shi et~al.(2015)Shi, Chang, and Cirac]{ShiChangCirac2015}
Tao Shi, Darrick~E. Chang, and J.~Ignacio Cirac.
\newblock Multiphoton-scattering theory and generalized master equations.
\newblock \emph{Physical Review A}, 92\penalty0 (5):\penalty0 053834, 2015.
\newblock \doi{10.1103/PhysRevA.92.053834}.

\bibitem[Crespi et~al.(2015)Crespi, Sansoni, Della~Valle, Ciamei, Ramponi,
  Sciarrino, Mataloni, Longhi, and Osellame]{Crespi2015}
Andrea Crespi, Linda Sansoni, Giuseppe Della~Valle, Alessio Ciamei, Roberta
  Ramponi, Fabio Sciarrino, Paolo Mataloni, Stefano Longhi, and Roberto
  Osellame.
\newblock Particle statistics affects quantum decay and {Fano} interference.
\newblock \emph{Physical Review Letters}, 114\penalty0 (9):\penalty0 090201,
  2015.
\newblock \doi{10.1103/PhysRevLett.114.090201}.

\bibitem[Ke et~al.(2019)Ke, Poshakinskiy, Lee, Kivshar, and Poddubny]{Ke2019}
Yongguan Ke, Alexander~V. Poshakinskiy, Chaohong Lee, Yuri~S. Kivshar, and
  Alexander~N. Poddubny.
\newblock Inelastic scattering of photon pairs in qubit arrays with subradiant
  states.
\newblock \emph{Physical Review Letters}, 123\penalty0 (25):\penalty0 253601,
  2019.
\newblock \doi{10.1103/PhysRevLett.123.253601}.

\bibitem[Zhang et~al.(2020)Zhang, Yu, and M{\o}lmer]{Zhang2020}
Yu-Xiang Zhang, Chuan Yu, and Klaus M{\o}lmer.
\newblock Subradiant bound dimer excited states of emitter chains coupled to a
  one-dimensional waveguide.
\newblock \emph{Physical Review Research}, 2\penalty0 (1):\penalty0 013173,
  2020.
\newblock \doi{10.1103/PhysRevResearch.2.013173}.

\bibitem[Poddubny(2020)]{Poddubny2020}
Alexander~N. Poddubny.
\newblock Quasiflat band enabling subradiant two-photon bound states.
\newblock \emph{Physical Review A}, 101\penalty0 (4):\penalty0 043845, 2020.
\newblock \doi{10.1103/PhysRevA.101.043845}.

\bibitem[Ke et~al.(2020)Ke, Zhong, Poshakinskiy, Kivshar, Poddubny, and
  Lee]{Ke2020}
Yongguan Ke, Janet Zhong, Alexander~V. Poshakinskiy, Yuri~S. Kivshar,
  Alexander~N. Poddubny, and Chaohong Lee.
\newblock Radiative topological biphoton states in modulated qubit arrays.
\newblock \emph{Physical Review Research}, 2\penalty0 (3):\penalty0 033190,
  2020.
\newblock \doi{10.1103/PhysRevResearch.2.033190}.

\bibitem[Te{\v{c}}er et~al.(2024)Te{\v{c}}er, Di~Liberto, Silvi, Montangero,
  Romanato, and Calaj{\'o}]{Tecer2024}
Matija Te{\v{c}}er, Marco Di~Liberto, Pietro Silvi, Simone Montangero, Filippo
  Romanato, and Giuseppe Calaj{\'o}.
\newblock Strongly interacting photons in two-dimensional waveguide {QED}.
\newblock \emph{Physical Review Letters}, 132\penalty0 (16):\penalty0 163602,
  2024.
\newblock \doi{10.1103/PhysRevLett.132.163602}.

\bibitem[Chang(2025)]{Chang2025NonMarkovian}
Yue Chang.
\newblock Non-{M}arkovian multiphoton chiral dynamics with giant systems.
\newblock \emph{Communications Physics}, 8:\penalty0 385, 2025.
\newblock \doi{10.1038/s42005-025-02293-w}.

\bibitem[Mirhosseini et~al.(2018)Mirhosseini, Kim, Ferreira, Kalaee, Sipahigil,
  Keller, and Painter]{Mirhosseini2018}
Mohammad Mirhosseini, Eunjong Kim, Vinicius~S. Ferreira, Mahmoud Kalaee, Alp
  Sipahigil, Andrew~J. Keller, and Oskar Painter.
\newblock Superconducting metamaterials for waveguide quantum electrodynamics.
\newblock \emph{Nature Communications}, 9:\penalty0 3706, 2018.
\newblock \doi{10.1038/s41467-018-06142-z}.

\bibitem[Vadiraj et~al.(2021)Vadiraj, Ask, McConkey, Nsanzineza, Sandbo~Chang,
  Kockum, and Wilson]{Vadiraj2021}
A.~M. Vadiraj, Andreas Ask, T.~G. McConkey, I.~Nsanzineza, C.~W. Sandbo~Chang,
  Anton~Frisk Kockum, and C.~M. Wilson.
\newblock Engineering the level structure of a giant artificial atom in
  waveguide quantum electrodynamics.
\newblock \emph{Physical Review A}, 103\penalty0 (2):\penalty0 023710, 2021.
\newblock \doi{10.1103/PhysRevA.103.023710}.

\bibitem[Sup(2026)]{SupplementalMaterial}
Supplemental material for ``two-photon bound states in the continuum: A no-go
  theorem and long-lived quasi-bound states'', 2026.
\newblock See the Supplemental Material for derivations of the no-go theorem
  for interaction-active two-photon BICs, the conditions for
  interaction-inactive two-photon BICs, the two-photon quasi-BIC of a giant
  Kerr cavity coupled to a waveguide with linear dispersion, and the reversed
  hierarchy between single- and two-photon decay rates.

\bibitem[Zhao and Wang(2020)]{ZhaoWang2020}
Wei Zhao and Zhihai Wang.
\newblock Single-photon scattering and bound states in an atom-waveguide system
  with two or multiple coupling points.
\newblock \emph{Physical Review A}, 101:\penalty0 053855, 2020.
\newblock \doi{10.1103/PhysRevA.101.053855}.

\bibitem[Guo et~al.(2020{\natexlab{a}})Guo, Wang, Purdy, and
  Taylor]{Guo2020BIC}
Shangjie Guo, Yidan Wang, Thomas Purdy, and Jacob Taylor.
\newblock Beyond spontaneous emission: Giant atom bounded in the continuum.
\newblock \emph{Physical Review A}, 102:\penalty0 033706, 2020{\natexlab{a}}.
\newblock \doi{10.1103/PhysRevA.102.033706}.

\bibitem[Guo et~al.(2020{\natexlab{b}})Guo, Kockum, Marquardt, and
  Johansson]{Guo2020Oscillating}
Lingzhen Guo, Anton~Frisk Kockum, Florian Marquardt, and G{\"o}ran Johansson.
\newblock Oscillating bound states for a giant atom.
\newblock \emph{Physical Review Research}, 2:\penalty0 043014,
  2020{\natexlab{b}}.
\newblock \doi{10.1103/PhysRevResearch.2.043014}.

\bibitem[Gu et~al.(2024)Gu, Li, Tian, Yi, and Li]{Gu2024}
Wenju Gu, Tao Li, Ye~Tian, Zhen Yi, and Gao-xiang Li.
\newblock Two-photon dynamics in non-{Markovian} waveguide {QED} with a giant
  atom.
\newblock \emph{Physical Review A}, 110:\penalty0 033707, 2024.
\newblock \doi{10.1103/PhysRevA.110.033707}.

\end{thebibliography}

\end{document}